\documentclass[a4paper,notitlepage,11pt]{article}

\usepackage[utf8]{inputenc}
\usepackage[T1]{fontenc} 
\usepackage[english]{babel}

\usepackage{amsmath,amssymb,amsthm,amstext,latexsym,eufrak,euscript}
\usepackage{subfigure,pstricks,pst-node,pst-coil}
\usepackage{dsfont}
\usepackage{stmaryrd}
\usepackage{cite}
\usepackage{graphicx}
\usepackage{algorithm2e}
\usepackage{listings}
\usepackage{epsfig}
\usepackage{url}

\usepackage{calc}
\usepackage{multicol}
\usepackage{pslatex}

\newcommand{\JFC}[1]{\begin{color}{green}\textit{}\end{color}}
\newcommand{\CG}[1]{\begin{color}{blue}\textit{}\end{color}}

\newtheorem*{xpl}{Running example}
\newtheorem{theorem}{Theorem}

\newcommand{\Nats}[0]{\ensuremath{\mathbb{N}}}

\newcommand{\Bool}[0]{\ensuremath{\mathds{B}}}

\usepackage[lmargin=2.5cm, rmargin=2.5cm,bottom=2.5cm,top=2.5cm]{geometry}

\author{J.-F. Couchot, P.-C. Heam, C. Guyeux, Q. Wang,  and J. M. Bahi\\
FEMTO-ST Institute, University of Franche-Comt\'{e}, France\\
College of Automation, Guangdong University of Technology, China\\
\{jean-francois.couchot,pierre-cyrille.heam,christophe.guyeux,\\jacques.bahi\}@univ-fcomte.fr, 
\\wangqianxue@gdut.edu.cn
}

\title {Traversing a $n$-cube without Balanced Hamiltonian Cycle \\
  to Generate Pseudorandom Numbers}

\begin{document}
\maketitle
\thispagestyle{empty} % Necessary in order for pagestyle{empty} to work
\pagestyle{empty}

%\keywords{Pseudorandom Number Generator, Theory of Chaos, Markov Matrices, Hamiltonian Paths, Mixing Time, Gray Codes, Statistical Tests}

\begin{abstract}
\JFC{Synthetiser}
This article presents a new  class
of Pseudorandom Number Generators. 
The generators are based on traversing a $n$-cube where  
a Balanced Hamiltonian Cycle has been removed. The construction of such generators is automatic for small number of bits, but remains an open problem when this 
number becomes large. A running example is used throughout the paper. Finally, first statistical experiments of these generators are presented, they show how efficient and promising the proposed approach seems.
\end{abstract}

\section{{Introduction}}
Many fields of research or applications like numerical simulations, 
stochastic optimization, or information security are highly dependent on
the use of fast and unbiased random number generators. Depending on the
targeted application, reproducibility must be either required, leading
to deterministic algorithms that produce numbers as close as possible
to random sequences, or refused, which implies to use an external
physical noise. The former are called pseudorandom number generators
(PRNGs) while the latter are designed by truly random number generators (TRNGs).
TRNGs are used for instance in cypher keys generation, or in hardware based
simulations or security devices. Such TRNGs are often based on a chaotic
physical signal, may be quantized depending on the application.
This quantization however raises the problem of the degradation of chaotic 
properties.

The use of PRNGs, for its part, is a necessity in a large variety 
of numerical simulations, in which
responses of devices under study must be compared using the same ``random''
stream. This reproducibility is required too for symmetric encryption like
one-time pad, as sender and receiver must share the same pad. However, in
that situation, security of the pseudorandom stream must be mathematically
proven: an attacker must not be able to computationally distinguish a
pseudorandom sequence generated by the considered PRNG with a really random
one. Such cryptographically secure pseudorandom number generators are
however only useful in cryptographic contexts, due to their slowness
resulting from their security.

Other kind of properties are desired for PRNGs used in numerical simulations
or in programs that embed a Monte-Carlo algorithm. In these situations,
required properties are speed and random-like profiles of the generated 
sequences. The fact that a given PRNG is unbiased and behaves as a white
noise is thus verified using batteries of statistical tests on a large amount
of pseudorandom numbers. Reputed and up-to-date batteries are
currently the NIST suite~\cite{Nist10}, and DieHARD~\cite{Marsaglia1996}.
 Finally, chaotic properties can be desired when
simulating a chaotic physical phenomenon or in hardware security, in 
which cryptographical proofs are not realizable.
In both truly and pseudorandom number generation, there is thus a need
to mathematically guarantee the presence of chaos, and to show that 
a post-treatment on a given secure and/or unbiased generator can be realized,
which adds chaos without deflating these desired properties.

This work takes place in this domain with the desire of automatically 
generating a large class of PRNGs with chaos and statistical properties.
In a sense, it is close to~\cite{BCGR11} where the authors shown that 
some Boolean maps may be embedded into an algorithm to provide a PRNG that has both
the theoretical Devaney's chaos property and the practical 
property of succeeding NIST statistical battery of tests.  
To achieve this, it has been proven in 
this article that it is sufficient   
for the iteration graph to be strongly connected,
and it is necessary and sufficient for its Markov probability matrix to be doubly stochastic.
However, they do not purpose conditions 
to provide such Boolean maps.
Admittedly, sufficient conditions
to retrieve Boolean maps whose graphs are 
strongly connected are given, but it remains to further filter those whose 
Markov matrix is doubly stochastic.
This approach suffers from delaying the second requirement to a final step
whereas this is a necessary condition. 
In this position article, we provide a completely new approach
to generate Boolean functions, whose Markov matrix is doubly stochastic and whose
graph of iterations is strongly connected. 
Furthermore the rate of convergence is always taken into consideration to provide 
PRNG with good statistical properties.

This research work is organized as follows.
It firstly recall some preliminaries that make the document self-contained (Section~\ref{sec:preliminaries}),
The next section (Section~\ref{sec:DSSC}) shows how the 
problem of finding some kind of matrices is moved into the graph theory.
Section~\ref{sec:hamiltonian} is the strongest contribution of this work.
It presents the main algorithm to generate Boolean maps with all the required properties and 
proves that such a construction is correct. 
Statistical evaluations are then summarized in Section~\ref{sec:exp}. 
Conclusive remarks, open problems, and perspectives are 
finally provided.

 \section{{Preliminaries}}\label{sec:preliminaries}

In what follows, we consider the Boolean algebra on the set 
$\Bool=\{0,1\}$ with the classical operators of conjunction '.', 
of disjunction '+', of negation '$\overline{~}$', and of 
disjunctive union $\oplus$. 
Let $n$ be a positive integer. A  {\emph{Boolean map} $f$ is 
a function from the Boolean domain 
 to itself 
such that 
$x=(x_1,\dots,x_n)$ maps to $f(x)=(f_1(x),\dots,f_n(x))$.
Functions are iterated as follows. 
At the $t^{th}$ iteration, only the $s_{t}-$th component is
``iterated'', where $s = \left(s_t\right)_{t \in \mathds{N}}$ is a sequence of indices taken in $\llbracket 1;n \rrbracket$ called ``strategy''. Formally,
let $F_f: \llbracket1;n\rrbracket\times \Bool^{n}$ to $\Bool^n$ be defined by
\[
F_f(i,x)=(x_1,\dots,x_{i-1},f_i(x),x_{i+1},\dots,x_n).
\]
Then, let $x^0\in\Bool^n$ be an initial configuration
and $s\in
\llbracket1;n\rrbracket^\Nats$ be a strategy, 
the dynamics are described by the recurrence
\begin{equation}\label{eq:asyn}
x^{t+1}=F_f(s_t,x^t).
\end{equation}

Let be given a Boolean map $f$. Its associated   
{\emph{iteration graph}}  $\Gamma(f)$ is the
directed graph such that  the set of vertices is
$\Bool^n$, and for all $x\in\Bool^n$ and $i\in \llbracket1;n\rrbracket$,
the graph $\Gamma(f)$ contains an arc from $x$ to $F_f(i,x)$.

It is easy to associate a Markov Matrix $M$ to such a graph $G(f)$
as follows: 
$M_{ij} = \frac{1}{n}$ if there is an edge from $i$ to $j$ in $\Gamma(f)$ and $i \neq j$;  $M_{ii} = 1 - \sum\limits_{j=1, j\neq i}^n M_{ij}$; and $M_{ij} = 0$ otherwise.

\begin{xpl}
Let us consider for instance $n=3$. Let 
$f^*: \Bool^3 \rightarrow \Bool^3$ be defined by
$f^*(x_1,x_2,x_3)=(x_2 \oplus x_3, x_1 \oplus \overline{x_3},\overline{x_3})$.
The iteration graph $\Gamma(f^*)$ of this function is given in 
Figure~\ref{fig:iteration:f*} and its Markov matrix is given 
in Figure~\ref{fig:markov:f*}.

\begin{figure}[ht]
\null\hfill
\subfigure[Iteration Graph $\Gamma(f^*)$ of the function $f^*$\label{fig:iteration:f*}]{
\begin{minipage}{0.3\textwidth}
\includegraphics[scale=0.5]{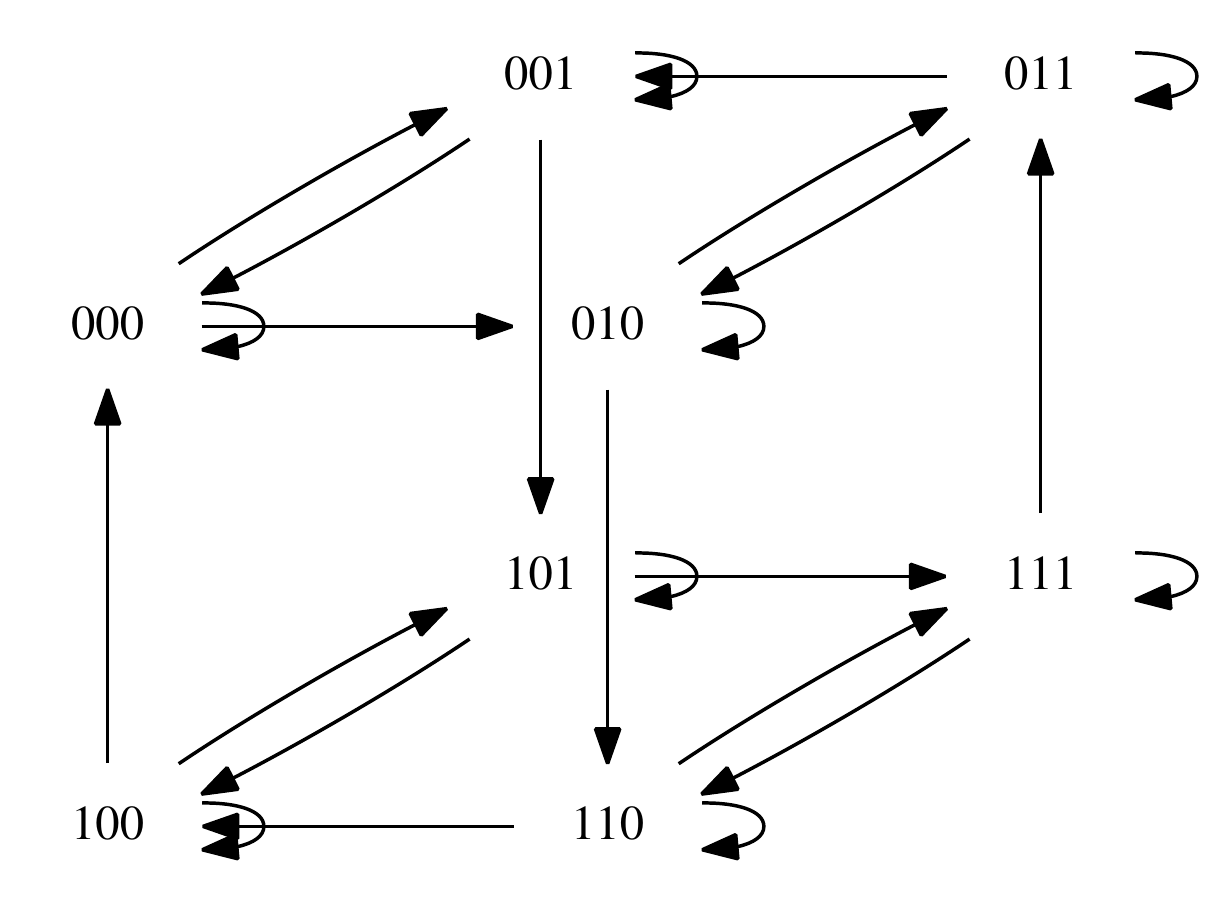}%
\end{minipage}
}
\hfill%
\subfigure[Markov matrix associated to the function $f^*$\label{fig:markov:f*}]{%
\begin{minipage}{0.3\textwidth}
$
\dfrac{1}{3}
\left(
\begin{array}{llllllll}
1 & 1 & 1 & 0 & 0 & 0 & 0 & 0 \\ 
1 & 1 & 0 & 0 & 0 & 1 & 0 & 0 \\ 
0 & 0 & 1 & 1 & 0 & 0 & 1 & 0 \\ 
0 & 1 & 1 & 1 & 0 & 0 & 0 & 0 \\ 
1 & 0 & 0 & 0 & 1 & 1 & 0 & 0 \\ 
0 & 0 & 0 & 0 & 1 & 1 & 0 & 1 \\ 
0 & 0 & 0 & 0 & 1 & 0 & 1 & 1 \\ 
0 & 0 & 0 & 1 & 0 & 0 & 1 & 1 
\end{array}
\right)
$
\end{minipage}
}%
\hfill\null
\caption{Representations of $f^*(x_1,x_2,x_3)=(x_2 \oplus x_3, x_1 \oplus \overline{x_3},\overline{x_3})$.}\label{fig1}
\end{figure}
\end{xpl}

The \emph{mixing time}~\cite{LevinPeresWilmer2006} is one of the usual metrics 
that gives how far the rows of a Markov matrix
converge to a specific distribution. 
It defines the smallest iteration number 
that is sufficient to obtain a deviation lesser than a given $\varepsilon$
for each rows of such kind of matrices.

Let us finally present the pseudorandom number generator $\chi_{\textit{14Secrypt}}$
which is based on random walks in $\Gamma(f)$. 
More precisely, let be given a Boolean map $f:\Bool^n \rightarrow \Bool^n$,
a PRNG \textit{Random},
an integer $b$ that corresponds to an awaited mixing time, and 
an initial configuration $x^0$. 
Starting from $x^0$, the algorithm repeats $b$ times 
a random choice of which edge to follow and traverses this edge.
The final configuration is thus outputted.
This PRNG is formalized in Algorithm~\ref{CI Algorithm} further denoted 
as $\chi_{\textit{14Secrypt}}$.

\begin{algorithm}[ht]
\begin{scriptsize}
%\JFC{Mettre ceci dans une boite flottante}
\KwIn{a function $f$, an iteration number $b$, an initial configuration $x^0$ ($n$ bits)}
\KwOut{a configuration $x$ ($n$ bits)}
$x\leftarrow x^0$\;
\For{$i=0,\dots,b-1$}
{
$s\leftarrow{\textit{Random}(n)}$\;
$x\leftarrow{F_f(s,x)}$\;
}
return $x$\;
\end{scriptsize}
\caption{Pseudo Code of the $\chi_{\textit{14Secrypt}}$ PRNG}
\label{CI Algorithm}
\end{algorithm}

% This PRNG is a particularized version of Algorithm given in~\cite{BCGR11}.
% Compared to this latter, the length of the random 
% walk of our algorithm is always constant (and is equal to $b$) whereas it 
% was given by a second PRNG in this latter.
% However, all the theoretical results that are given in~\cite{BCGR11} remain
% true since the proofs do not rely on this fact. 
% Let us recall the following theorem.

Let $f: \Bool^{n} \rightarrow \Bool^{n}$.
It has been shown~\cite[Th. 4, p. 135]{BCGR11}} that 
if its iteration graph is strongly connected, then 
the output of $\chi_{\textit{14Secrypt}}$ follows 
a law that tends to the uniform distribution 
if and only if its Markov matrix is a doubly stochastic matrix.

The next  section presents  an efficient method to
generate Boolean functions
with Doubly Stochastic matrix and Strongly Connected iteration graph,
further (abusively) denoted as DSSC matrix.

\section{{Generation of DSSC Matrices}}\label{sec:DSSC}
Finding DSSC matrices can be theoretically handled by  
Constraint Logic Programming on Finite Domains (CLPFD):
all the variables range into finite integer domains with sum and 
product operations. 
However, this approach suffers from not being efficient enough for
large $n$ due to a \emph{generate and test} pattern.

Intuitivelly, considering the $n$-cube and 
removing one outgoing edge and
one ongoing edge for each node should be a practical answer 
to the DSSC matrix finding problem. 
Moreover, the previous wish of exaclty removing exactly one outgoing 
and one ongoing edge for each node is solved by removing a 
Hamiltonian cycle in the $n$-cube. The next section details this step.

\begin{xpl}
The iteration graph of $f^*$ 
(given in Figure~\ref{fig:iteration:f*}) 
is the $3$-cube in which the Hamiltonian cycle 
$000,100,101,001,011,111,110,010,000$ 
has been removed.
\end{xpl}

\section{{Removing Hamiltonian Cycles}}\label{sec:hamiltonian}
The first theoretical section (Section~\ref{sub:removing:theory})
shows that this approach produces DSSC matrix,
as wished.
The motivation to focus on balanced Gray code is then given in Sec.~\ref{sub:gray}.
We end this section by giving some discussion about practical aspeccts 
of an existing 
algorithm that aims at computing such codes (Section~\ref{sub:algo:gray}).

\subsection{Theoretical Aspects of Removing Hamiltonian Cycles}
\label{sub:removing:theory}
We first have the following result on stochastic matrix and $n$-cube without
Hamiltonian cycle.

\begin{theorem}
The Markov Matrix $M$ resulting from the $n$-cube in
which an Hamiltonian 
cycle is removed, is doubly stochastic.
\end{theorem}

% \begin{proof}
% An Hamiltonian cycle visits each vertex exactly once. 
% For each vertex $v$ in the $n$-cube,
% one ongoing edge $(o,v)$ and one outgoing edge $(v,e)$ 
% are thus removed.

% Let us consider the Markov matrix $M$ of the $n$-cube. 
% It is obviously doubly stochastic.
% Since we exactly remove one outgoing edge, the value of $M_{ve}$ decreases
% from $\frac{1}{n}$  to 0 and
% $M_{vv}$ is $\frac{1}{n}$. The $M$ matrix is stochastic again.
% Similarly for ongoing edge, since one ongoing edge is dropped for each 
% vertex, the value of $M_{ov}$ decreases
% from $\frac{1}{n}$  to $0$. Moreover, since $M_{vv}$ is $\frac{1}{n}$,
% the sum of values in column $v$ is $1$, and $M$ is doubly stochastic. 
% \end{proof}

The proof is left as an exercise for the reader.
The following result states that the $n$-cube without one
Hamiltonian cycle 
has the awaited property with regard to the connectivity.

\begin{theorem}
The iteration graph issued from the $n$-cube where an Hamiltonian 
cycle is removed is strongly connected.
\end{theorem}

% \begin{proof}
% We consider the reverse cycle $r$ of the Hamiltonian cycle $c$.
% There is no edge that belongs to both $r$ and $c$: otherwise $c$
% would contain one vertex twice. Thus, no edges of $r$ has been removed.
% The cycle $r$ is obviously an Hamiltonian cycle and contains all the nodes.
% Any node of the $n$-cube where $c$ has been removed  can thus reach any node.  
% The  iteration graph is thus strongly connected.  
% \end{proof}
Again, the proof is left as an exercise for the reader.
Removing an Hamiltonian cycle in the $n$-cube solves thus the DSSC constraint.
We are then left to focus on the generation of Hamiltonian cycles in the 
$n$-cube. Such a problem is equivalent to find cyclic Gray codes, \textit{i.e.},
to find a sequence of $2^n$ codewords ($n$-bits strings) 
where two successive elements differ in only one bit position and
and where the last codeword   
differs in only one bit position from the first one.
The next section is dedicated to these codes.

\subsection{Linking to Cyclic (Totally) Balanced Gray Codes}\label{sub:gray}
% Many research works~\cite{journals/combinatorics/BhatS96,VanSup04,journals/combinatorics/FlahiveB07} 
% have addressed the subject of finding Gray  codes.

Let $n$ be a given integer. As far as we know, 
the exact number of Gray codes in $\Bool^n$ is not known but 
a lower bound, $\left(\frac{n*\log2}{e \log \log n}*(1 - o(1))\right)^{2^n}$
has been given in~\cite{Feder:2009:NTB:1496397.1496515}.
For example, when $n$ is $6$, such a number is larger than $10^{13}$.
To avoid this combinatorial explosion, we want to
restrict the generation to any Gray code 
such that the induced graph of iteration $\Gamma(f)$ is 
``uniform''. In other words, if we count in $\Gamma(f)$ 
the number of edges that modify the bit $i$, for $1 \le i \le n$,
all these values have to be close to each other.
Such an approach is equivalent to restrict the search of cyclic Gray codes
which are uniform too.

This notion can be formalized as follows. Let  $L = w_1, w_2, \dots, w_{2^n}$ be the sequence 
of a $n$-bits cyclic Gray code. Let $S = s_1, s_2, \dots, s_{2^n}$ be the 
transition sequence where $s_i$, $1 \le i \le 2^n$ indicates which bit position changes between 
codewords at index $i$ and $i+1$ modulo $2^n$. 
Let $\textit{TC}_n : \{1,\dots, n\} \rightarrow \{0, \ldots, 2^n\}$ the \emph{transition count} function
that counts the number of times $i$ occurs in $S$, \textit{i.e.}, the number of times 
the bit $i$ has been switched in $L$.   
The Gray code is \emph{totally balanced} if $\textit{TC}_n$ is constant (and equal to $\frac{2^n}{n}$).
It is \emph{balanced} if for any two bit indices $i$ and $j$, 
$|\textit{TC}_n(i) - \textit{TC}_n(j)| \le  2$.
   
\begin{xpl}
Let $L^*=000,100,101,001,011,111,110,010$ be the Gray code that corresponds to 
the Hamiltonian cycle that has been removed in $f^*$.
Its transition sequence is $S=3,1,3,2,3,1,3,2$ and its transition count function is 
$\textit{TC}_3(1)= \textit{TC}_3(2)=2$ and  $\textit{TC}_3(3)=4$. Such a Gray code is balanced. 

Let now  
$L^4=0000, 0010, 0110, 1110, 1111, 0111, 0011, 0001, 0101,$
$0100, 1100, 1101, 1001, 1011, 1010, 1000$
be a cyclic Gray code. Since $S=2,3,4,1,4,3,2,3,1,4,1,3,2,1,2,4$,
its transition count $\textit{TC}_4$ is equal to 4 everywhere and this code
is thus totally balanced.
% On the contrary, for the standard $4$-bits Gray code  
% $L^{\textit{st}}=0000,0001,0011,0010,0110,0111,0101,0100,1100,$
% $1101,1111,1110,1010,1011,1001,1000$,
% we have $\textit{TC}_4(1)=8$ $\textit{TC}_4(2)=4$ $\textit{TC}_4(3)=\textit{TC}_4(4)=2$ and
% the code is neither balanced nor totally balanced.
\end{xpl}

\subsection{Induction-Based Generation of  Balanced Gray Codes}\label{sub:algo:gray} 

The article~\cite{ZanSup04} proposed the ``Construction B'' algorithm to produce Balanced Gray Codes. This method inductively constructs $n$-bits Gray code given a $n-2$-bit Gray code.  The authors have proven  that 
$S_{n}$ is transition sequence of a cyclic $n$-bits Gray code 
if $S_{n-2}$ is. It starts with the transition sequence $S_{n-2}$ of such code 
and the following first step:

\emph{
Let $l$ be an even positive integer. Find 
$u_1, u_2, \dots , u_{l-2}, v$ (maybe empty) subsequences of $S_{n-2}$ 
such that $S_{n-2}$ is the concatenation of 
$
s_{i_1}, u_0, s_{i_2}, u_1, s_{i_3}, u_2, . . . , s_{i_l-1}, u_{l-2}, s_{i_l}, v
$
where $i_1 = 1$, $i_2 = 2$, and $u_0 = \emptyset$ (the empty sequence).
}

However, this first step is not  constructive: 
it does not precises how to select the subsequences which ensures that 
yielded Gray code is balanced.

Let us now  evaluate the number of subsequences $u$  
than can be produced. Since $s_{i_1}$ and $s_{i_2}$ are well
defined, we have to chose the $l-2$ elements of $s_3,s_4,\dots,s_{2^{n-2}}$
that become $s_{i_3},\dots, s_{i_l}$. Let $l = 2l'$.
There are thus 
$
\#_n = \sum_{l'=1}^{2^{n-3}} {2^{n-2}-2 \choose 2l'-2}
$
distinct subsequences $u$.
Numerical values of $\#_n$ are given in table~\ref{table:combinations}.
Even for small values of $n$, it is not reasonable to hope to evaluate the whole 
set of subsequences.
\begin{table}[ht]
\begin{center}
\begin{tabular}{|l|l|l|l|l|l|l|l|l|l|}
\hline
$n$ & 4&  5 & 6 & 7 & 8  \\  
\hline
$\#_n$ & 
1& 31 & 8191 & 5.3e8 & 2.3e18 \\
\hline
$\#'_n$ & 
1 & 15 & 3003 & 1.4e8 & 4.5e17  \\
\hline
\end{tabular}
\end{center}
\caption{Number of distinct $u$ subsequences.}\label{table:combinations}
\end{table}

However, it is  shown in the article that $\textit{TC}_n(n-1)$ and $\textit{TC}_n(n)$ are
equal to $l$. Since this step aims at generating (totally) balanced Gray codes,  
we have set $l$ to be the largest even integer less or equal than $\frac{2^n}{n}$.
This improvement allows to reduce the number of subsequences to study.
Examples of such cardinalities are given in Table~\ref{table:combinations} and are referred as 
$\#'_n$.

Finally, the table~\ref{table:nbFunc} gives the number of non-equivalent functions issued from 
(totally) balanced Gray codes that can be generated
with the approach presented in this article with respect to the number of bits. In other words, it corresponds to the size of the class of generators that can be produced. Notice that when $n$ is 7 and 8, we only give lower bounds for 
2.5E5 distinct choices for the $u$ subsequence.

\begin{table}[ht]
\begin{center}
\begin{tabular}{|l|c|c|c|c|c|}
\hline
$n$              &  4 & 5 & 6    & 7 & 8 \\
\hline
nb. of functions &  1 & 2 & 1332 & > 2300  & > 4500   \\
\hline
\end{tabular}
\end{center}
\caption{Number of Generators w.r.t. the number of bits.}
\label{table:nbFunc}
\end{table}

\section{{Experiments}}\label{sec:exp}
We have directly implemented the algorithm given in Figure~\ref{CI Algorithm}.
For function $f$ and our experiments $b$ is set
with the value given in the fourth column of Table~\ref{table:nc}.
% \begin{algorithm}[ht]
% %\begin{scriptsize}
% \KwIn{a function $f$, an iteration number $b$, an initial configuration $x^0$ ($n$ bits)}
% \KwOut{a configuration $x$ ($n$ bits)}
% $x\leftarrow x^0$\;
% \For{$i=0,\dots,b-1$}
% {
% $s\leftarrow{\textit{Random}~mod~n}$\;
% $x\leftarrow{(x-(x\&(1<<s))+f(x)\&(1<<s))}$\;
% }
% return $x$\;
% %\end{scriptsize}
% \caption{Pseudo Code of the $\chi_{\textit{14Secrypt}}$ PRNG}
% \label{CI}
% \end{algorithm}

\begin{table*}[t]
\begin{center}
\begin{scriptsize}
\begin{tabular}{|c|l|l|c|}
\hline
Function $f$ & $f(x)$, for $x$ in $(0,1,2,\hdots,2^n-1)$ & $n$ & $b$ \\ 
%%%%% n= 4
\hline
$\textcircled{a}$&[13,10,9,14,3,11,1,12,15,4,7,5,2,6,0,8]&4&32\\
\hline
%%%%% n= 5
$\textcircled{b}$& [29, 22, 21, 30, 19, 27, 24, 28, 7, 20, 5, 4, 23, 26, 25, 17, 31, 12, 15, 8, 10, 14, 13, 9, 3, 2, 1, 6, 11, 18, 0, 16] & 5 & 41 \\
%%%%% n= 6
\hline
&[55, 60, 45, 56, 43, 62, 61, 40, 53, 50, 52, 36, 59, 34, 57, 49, 15, 14, 47, 46, 11, 58, 33, 44, 7, 54, 39, 37, 51, 2, 32, 48, 
&&\\
$\textcircled{c}$&
63, 26, 25, 30, 19, 27, 17, 28, 31, 20, 23, 21, 18, 22, 16, 24, 13, 12, 29, 8, 10, 42, 41, 0, 5, 38, 4, 6, 35, 3, 9, 1]
&6&49\\
%%%%% n= 7
\hline
&[111,94,93,116,122,114,125,88,87,126,119,84,123,98,81,120,109,78,105,110,99,107,104,108,101,
&&\\
& 70,117,96,103,102,113,64,79,30,95,124,83,91,121,24,85,118,69,20,115,90,17,112,77,76,73,12,74, &&\\
$\textcircled{d}$&106,72,8,7,6,71,100,75,82,97,0,127,54,57,62,51,59,56,48,53,38,37,60,55,58,33,49,63,44,47,40,42,&7 & 63\\
&46,45,41,35,34,39,52,43,50,32,36,29,28,61,92,26,18,89,25,19,86,23,4,27,2,16,80,31,10,15,14,3,11,&&\\
&13,9,5,22,21,68,67,66,65,1]
&&\\
%%%%%n=8
\hline
&[223,250,249,254,187,234,241,252,183,230,229,180,227,178,240,248,237,236,253,172,251,238,201,&&\\
&
  224,247,166,165,244,
163,242,161,225,215,220,205,216,218,222,221,208,213,210,135,196,199,132,
&&\\
&
194,130,129,200,159,186,185,190,59,170,177,188,191,246,245,52,243,50,176,184,173,46,189,174,
&&\\
&
235,42,233,232,231,38,37,228,35,226,33,168,151,156,141,152,
154,158,157,144,149,146,148,150,
&&\\
$\textcircled{e}$&
155,147,153,145,175,14,143,204,11,202,169,8,7,198,197,4,195,2,1,192,255,124,109,120,107,126,
&8&75\\
&
125,112,103,114,116,100,123,98,121,113,79,106,111,110,75,122,97,108,71,118,117,68,115,66,96,
&&\\
&
104,127,90,89,94,83,91,81,92,95,84,87,85,82,86,80,88,77,76,93,72,74,78,105,64,69,102,101,70,99,
&&\\
&
67,73,65,55,60,45,56,51,62,61,48,119,182,181,53,179,54,57,49,15,44,47,40,171,58,9,32,167,6,5,
&&\\
&
164,3,162,41,160,63,26,25,30,19,27,17,28,31,20,23,21,18,22,16,24,13,10,29,140,43,138,137,12,39,
&&\\
&
134,133,36,131,34,0,128]&&\\
\hline

\end{tabular}
\end{scriptsize}
\end{center}
\caption{Functions with DSCC Matrix and smallest MT\label{table:nc}}
\end{table*}

For each number $n=4,5,6,7,8$ of bits, we have generated 
the functions according the method 
given in Section~\ref{sub:algo:gray} .
For each $n$, we have then restricted this evaluation to the function 
whose Markov Matrix has the smallest mixing time. Such functions are 
given in Table~\ref{table:nc}.
In this table, let us consider for instance 
the function $\textcircled{a}$ from $\Bool^4$ to $\Bool^4$
defined by the following images : 
$[13, 10, 9, 14, 3, 11, 1, 12, 15, 4, 7, 5, 2, 6, 0, 8]$.
In other words,  the image of $3~(0011)$ by $\textcircled{a}$ is $14~(1110)$: it is obtained as  the  binary  value  of  the  fourth element  in  the  second  list
(namely~14).  

Experiments have shown that all the generators pass the NIST and the DieHARD batteries of tests.

% \subsection{TestU01}
% \label{Subsec:TestU01}

% TestU01~\cite{Simard07testu01:a} is a software library that 
% provides general implementations of the classical statistical tests for 
% random number generators, as well as several others proposed in the 
% literature, and some original ones. This library is currently the
% most reputed and stringent one for testing the randomness
% profile of a given sequence. TestU01 encompasses the NIST and DieHARD
% tests suites with 2 batteries specific to hardware based generators 
% (namely, Rabbit and Alphabit). Its three core batteries of tests are
% however SmallCrush, Crush, and BigCrush, classified according to
% their difficulty.

% To date, we can claim after experiments that $\textcircled{a}$
% generator is able to pass the 15 tests embedded into the 
% SmallCrush battery and it succedded too to pass the 144 tests of 
% the Crush one. BigCrush results on $\textcircled{a}$ are
% expected soon, while TestU01 has been launched too on generators
% having the other iteration functions detailed in this article.

\section{{Conclusion}}\label{sec:conclusion}
This article has presented a method to compute a large class of truly chaotic PRNGs.
First experiments through the batteries of NIST, and DieHard 
have shown that the statistical properties are almost established for $n=4,5,6,7,8$.
The iterated map inside the generator is built by removing 
from a $n$-cube an Hamiltonian path that corresponds to a (totally) balanced Gray code.  
The number of balanced gray code is large and each of them can be 
considered as a key of the PRNG.
However, many problems still remain open, most important ones being listed 
thereafter. 

The first one involves the function to iterate. Producing a DSSC matrix is indeed necessary and sufficient
but is not linked with the convergence rate to the uniform distribution. 
To solve this problem, we have proposed to remove from the $n$-cube an Hamiltonian path that 
is a (totally) balanced Gray code. We do not have proven that this 
proposal is the one that minimizes the  mixing time. This optimization task is an open problem we plan to study.

Secondly, the approach depends on finding (totally) balanced Gray 
codes. Even if such codes exist for all even numbers, there is no constructive method to built them
when $n$ is large, as far as we know. These two open problems will
be investigated in a future work.

\bibliographystyle{plain}
{\small
\bibliography{markov}}
\end{document}